\documentstyle[epsf]{mn}

\newif\ifAMStwofonts
\AMStwofontstrue

\newcommand{\be}{\begin{equation}}
\newcommand{\ee}{\end{equation}}
\ifoldfss
  \newcommand{\rmn}[1] {{\rm #1}}

  \ifCUPmtlplainloaded \else
    \NewTextAlphabet{textbfit} {cmbxti10} {}
    \NewTextAlphabet{textbfss} {cmssbx10} {}
    \NewMathAlphabet{mathbfit} {cmbxti10} {} % for math mode
    \NewMathAlphabet{mathbfss} {cmssbx10} {} %  "   "    "
  \fi
  \ifAMStwofonts
    \ifCUPmtlplainloaded \else
      \NewSymbolFont{upmath} {eurm10}
      \NewSymbolFont{AMSa} {msam10}
      \NewMathSymbol{\upi}     {0}{upmath}{19}
      \NewMathSymbol{\umu}     {0}{upmath}{16}
      \NewMathSymbol{\upartial}{0}{upmath}{40}
      \NewMathSymbol{\leqslant}{3}{AMSa}{36}
      \NewMathSymbol{\geqslant}{3}{AMSa}{3E}

      \let\geq=\geqslant 
    \fi
  \fi
\fi % End of OFSS

\ifnfssone
  \newmathalphabet{\mathit}
  \addtoversion{normal}{\mathit}{cmr}{m}{it}
  \addtoversion{bold}{\mathit}{cmr}{bx}{it}
  \newcommand{\rmn}[1] {\mathrm{#1}}

  \newmathalphabet{\mathbfit} % math mode version of \textbfit{..}
  \addtoversion{normal}{\mathbfit}{cmr}{bx}{it}
  \addtoversion{bold}{\mathbfit}{cmr}{bx}{it}
  \newmathalphabet{\mathbfss} % math mode version of \textbfss{..}
  \addtoversion{normal}{\mathbfss}{cmss}{bx}{n}
  \addtoversion{bold}{\mathbfss}{cmss}{bx}{n}
  \ifAMStwofonts
    \ifCUPmtlplainloaded \else
      \UseAMStwoboldmath
      \makeatletter
      \new@mathgroup\upmath@group
      \define@mathgroup\mv@normal\upmath@group{eur}{m}{n}
      \define@mathgroup\mv@bold\upmath@group{eur}{b}{n}
      \edef\UPM{\hexnumber\upmath@group}
      \new@mathgroup\amsa@group
      \define@mathgroup\mv@normal\amsa@group{msa}{m}{n}
      \define@mathgroup\mv@bold\amsa@group{msa}{m}{n}
      \edef\AMSa{\hexnumber\amsa@group}
      \makeatother
      \mathchardef\upi="0\UPM19
      \mathchardef\umu="0\UPM16
      \mathchardef\upartial="0\UPM40
      \mathchardef\leqslant="3\AMSa36
      \mathchardef\geqslant="3\AMSa3E

      \let\geq=\geqslant 
    \fi
  \fi
\fi % End of NFSS release 1

\ifnfsstwo
  \newcommand{\rmn}[1] {\mathrm{#1}}

  \DeclareMathAlphabet{\mathbfit}{OT1}{cmr}{bx}{it}
  \SetMathAlphabet\mathbfit{bold}{OT1}{cmr}{bx}{it}
  \DeclareMathAlphabet{\mathbfss}{OT1}{cmss}{bx}{n}
  \SetMathAlphabet\mathbfss{bold}{OT1}{cmss}{bx}{n}
  \ifAMStwofonts
    \ifCUPmtlplainloaded \else
      \DeclareSymbolFont{UPM}{U}{eur}{m}{n}
      \SetSymbolFont{UPM}{bold}{U}{eur}{b}{n}
      \DeclareSymbolFont{AMSa}{U}{msa}{m}{n}
      \DeclareMathSymbol{\upi}{0}{UPM}{"19}
      \DeclareMathSymbol{\umu}{0}{UPM}{"16}
      \DeclareMathSymbol{\upartial}{0}{UPM}{"40}
      \DeclareMathSymbol{\leqslant}{3}{AMSa}{"36}
      \DeclareMathSymbol{\geqslant}{3}{AMSa}{"3E}

      \let\geq=\geqslant 
    \fi
  \fi
\fi % End of NFSS release 2

\ifCUPmtlplainloaded \else
  \ifAMStwofonts \else % If no AMS fonts
    \def\upi{\pi}
    \def\umu{\mu}
    \def\upartial{\partial}
  \fi
\fi

\begin{document}

\title[Evolution of Cosmological Perturbations]{Validity of the
`Conservation Law' in the Evolution of Cosmological Perturbations}

\author[M. G\"{o}tz]{Martin G\"{o}tz\\
        Harvard-Smithsonian Center for Astrophysics, 60 Garden St., MS-20,
        Cambridge, MA 02138, USA and\\
        Physics Department, Brown University, Box 1843, Providence, RI 02912,
        USA, e-mail: {\tt mgotz@cfa.harvard.edu}}
\maketitle

\begin{abstract}
It is shown that the `conservation law' for metric
fluctuations with long wavelengths is indeed applicable for growing modes of
perturbations, which are of interest in cosmology, in spite of a recent
criticism \cite{gr}. This is demonstrated
both by general arguments, and also by presenting an explicitly
solvable toy model for the evolution of metric perturbations during
inflation.
\end{abstract}

\section{Introduction}

The gauge-invariant theory of cosmological perturbations (see
Mukhanov, Feldman \& Brandenberger \shortcite{review}
for a review) has been used successfully to calculate the origin and
evolution of metric and density perturbations in the early universe. The
most beautiful aspect of this approach is that the evolution of metric
perturbations with wavelengths larger than the Hubble radius, where the
growth in amplitude during inflation takes places, is characterized by a
simple conservation law
\be
\zeta = \frac{2}{3} \frac{\dot{\Phi}/H+\Phi}{1+w} + \Phi = \rmn{const.},
\label{con}
\ee
where $\Phi$ denotes the gauge-invariant generalized Newtonian potential for
scalar metric perturbations, $w=p/\epsilon$ the equation of state of the
matter content, and $H$ the Hubble parameter (Equ.~(5.23) and (6.58) in
Mukhanov et al. \shortcite{review}).
It should be noted though that this is valid only for
perturbations in a universe with a spatially flat background metric (i.e.
$K=0$), which was not stated after Equ.~(5.23) in
Mukhanov et al. \shortcite{review}.
Therefore only the case $K=0$ will be considered in this paper.

The growth of fluctuations in a
universe with a de Sitter-phase is apparent from (\ref{con}), since
$w \sim -1$ during inflation leads to a small denominator for $\zeta$ such
that the amplitude of $\Phi$ has to increase considerably in order to keep
$\zeta$ constant as the universe exits inflation into a radiation
($w = 1/3$) or matter dominated phase ($w = 0$).

Recently, there has been a criticism by Grishchuk \shortcite{gr} that
the conservation of $\zeta$ is derived from an incorrect equation of motion
for the cosmological perturbations and that it is always zero in the limit
of long wavelengths, in which case
no statement on the evolution of scalar metric perturbations would be
possible, as $\Phi$ could be multiplied by any arbitrary factor without
affecting the value of $\zeta$.

In this paper we will show that the equation used in deriving that $\zeta$
is constant in the long-wavelength limit is indeed the correct one, and that
furthermore $\zeta$ always takes non-zero values
for growing modes and can become zero only for decaying ones. This will
be exemplified by explicitly constructing a toy model for which the
evolution of metric perturbations during inflation can be determined
analytically. This is different from previously proposed toy models for that
purpose \cite{ca} in which the equation for the metric perturbations had to
be integrated numerically.
Hence we show that the conservation law (\ref{con}) is indeed a useful
relation to describe the evolution of metric perturbations at large
wavelengths. Note that earlier criticism by Grishchuk \shortcite{gr1}
has been refuted by Deruelle \& Mukhanov \shortcite{dm} and by Caldwell
\shortcite{ca}.

\section{General Remarks}

The basic equation governing the time dependence of the metric fluctuations
$\Phi$ (Equ.~(5.22) and (6.49) in Mukhanov et al. \shortcite{review})
is a linear combination
of the time-time and space-space components of the linearized Einstein
perturbation equations expressed in terms of gauge-invariant variables. For
the case of a universe with a single scalar field ($\phi$ is its unperturbed
background value) this equation reads
\be
\Phi^{\prime\prime} + 2 \left(\frac{a}{\phi^{\prime}}\right)^{\prime}
\frac{\phi^{\prime}}{a} \Phi^{\prime} + \bmath{k}^2 \Phi + 2 \phi^{\prime}
\left(\frac{\mathcal H}{\phi^{\prime}}\right)^{\prime}\Phi = 0
\label{phiequ}
\ee
for a Fourier mode with wavenumber $\bmath{k}$, where primes denote
differentiation with respect to conformal time, $a$ the scale factor, and
${\mathcal H} = a^{\prime} / a$. The conservation law (\ref{con}) follows from
this with some simple algebra.

Alternatively, fluctuations can be described
by a linear combination of the metric perturbations $\Phi$
and the gauge-invariant scalar field perturbations $\delta\phi$
\[
v = a \left( \delta\phi + \frac{\phi^{\prime}}{\mathcal H}\Phi \right)
\]
(Equ.~(10.71) in Mukhanov et al. \shortcite{review};
$v$ is called $\mu$ and $\nu$ in Grishchuk \shortcite{gr})
which arises in the derivation of the Lagrangian for the Einstein
perturbation equations. Varying the corresponding action gives the equation
of motion
\be
v^{\prime\prime} + \bmath{k}^2 v - \frac{z^{\prime\prime}}{z}v = 0
\label{vequ}
\ee
for that variable, where $z=a \phi^{\prime} / {\mathcal H}$. By taking
advantage of the equation for the evolution of the scalar field background
and the constraint equation for $\Phi$, the
relation between the two gauge-invariant variables $\Phi$ and $v$ can be
brought in the form
\be
-\bmath{k}^2 \Phi = 4\upi G \frac{\phi^{\prime 2}}{\mathcal H}
\left(\frac{v}{z}\right)^{\prime}
\label{relation}
\ee
(Equ.~(13.7) in Mukhanov et al. \shortcite{review}) with
Newton's constant $G$.

We expect that both Equ.~(\ref{phiequ}) and Equ.~(\ref{vequ})
should yield the same physical results for the evolution of cosmological
perturbations, i.e. it should be possible to go back and forth between
these two different ways to formulate the problem. But as we substitute
relation (\ref{relation}) between the two gauge-invariant variables in
(\ref{phiequ}) in an attempt to recover (\ref{vequ}) we actually end up with
\be
\left( v^{\prime\prime} + \bmath{k}^2 v - \frac{z^{\prime\prime}}{z}v
\right)^{\prime} - \frac{z^{\prime}}{z}
\left( v^{\prime\prime} + \bmath{k}^2 v - \frac{z^{\prime\prime}}{z}v
\right) = 0,
\label{vequ1}
\ee
an equation which admits more solutions than (\ref{vequ}), from which
Grishchuk \shortcite{gr} argues that the equation for the metric perturbations
$\Phi$ is the wrong one to begin with.

Here I would like to argue that (\ref{phiequ}) and (\ref{vequ}) are still
equivalent. The problem is that in order to substitute (\ref{relation}) we
have to multiply the equation for $\Phi$ by $\bmath{k}^2$ (or $\nabla^2$ in
position space), thus it is no surprise that we find more solutions than
we need. For that reason (\ref{vequ1}) is {\it not} equivalent to
(\ref{phiequ}), but to another equation
\be
\Phi^{\prime\prime} + 2 \left(\frac{a}{\phi^{\prime}}\right)^{\prime}
\frac{\phi^{\prime}}{a} \Phi^{\prime} + \bmath{k}^2 \Phi + 2 \phi^{\prime}
\left(\frac{\mathcal H}{\phi^{\prime}}\right)^{\prime}\Phi = f(\bmath{k})
\label{phiequ1}
\ee
where the right-hand side is replaced by an arbitrary, possibly
time-dependent function subject only to the condition
$\bmath{k}^2 f(\bmath{k}) = 0$, e.g. a $\delta$-function. So we need to check
for each of the three basic solutions of (\ref{vequ1}) whether their
corresponding $\Phi$-modes give a zero or non-zero result on the right-hand
side of (\ref{phiequ1}) and disregard the latter ones.

It is easy to determine that the full set of solutions of (\ref{vequ1}) is
\[
v(\bmath{k},\eta) = C_1 v_1(\bmath{k},\eta) + C_2 v_2(\bmath{k},\eta)
+ C_3 \frac{z}{\bmath{k}^2}
\]
where $v_1$ and $v_2$ are the two linearly independent solutions of the
original equation (\ref{vequ}) for $v$ and $\eta$ denotes conformal time.
Exploiting that $v_1$ and $v_2$
satisfy (\ref{vequ}) we can rewrite the relation between $\Phi$ and $v$ as
\be
\Phi_i = \frac{4\upi G{\mathcal H}}{a^2} \int v_i z \, {\rmn{d}} \eta
+ h(\bmath{k}) , \; i=1,2
\label{relation1}
\ee
where $h(\bmath{k})$ is another arbitrary,
potentially time-dependent function satisfying $\bmath{k}^2 h(\bmath{k}) = 0$.
Using this relation in (\ref{phiequ1}) we see that
these two modes for $\Phi$ do indeed give zero on the right-hand side
as long as $h(0)$ is zero or itself a solution of the original
equation (\ref{phiequ}) for $\Phi$ for $\bmath{k}=0$, which means it has to
be a linear combination of $\Phi_1$ and $\Phi_2$ in that case. (For
$\bmath{k} \neq 0$ we already have $h(\bmath{k})=0$.) Since it does not matter
which linear combination we call the basic solutions we can set
$h(\bmath{k}) = 0$ in (\ref{relation1}) without loss of generality. For the
third mode $z / \bmath{k}^2$ of $v$ we have to use the original relation
(\ref{relation}) to obtain a third mode $\Phi_3  = h(\bmath{k})$ for $\Phi$,
where again we have some function with $\bmath{k}^2 h(\bmath{k}) = 0$.
This mode will, in general, produce a non-zero
result on the right-hand side of (\ref{phiequ1}), unless, like before, it is
a linear combination of the other two modes for $\bmath{k}=0$ when
$h(\bmath{k})$ might be non-zero, in which case it will not give any new
information.

Thus it makes no difference whether we describe cosmological
perturbations in terms of metric perturbations $\Phi$ alone, or with the
combined gauge-invariant variable $v$, as we can map the solutions of
(\ref{phiequ}) and (\ref{vequ}) into each other with relation
(\ref{relation1}) (and $h(\bmath{k}) = 0$). If we use this, we find after
some algebra that the conserved quantity in terms of $v$ is simply
\be
\zeta = \frac{v}{z},
\label{zeta1}
\ee
a result which is valid for any wavenumber $\bmath{k}$. As for long wavelengths
$v=z$ is one obvious solution to (\ref{vequ}) we see immediately that there
exist modes for which $\zeta$ is constant and non-zero in that limit. The
problem with the argument for $\zeta = 0$ in Grishchuk \shortcite{gr} is
that the limit $\bmath{k} \rightarrow 0$ is not taken consistently,
otherwise the author would
have recovered (\ref{zeta1}) instead of $\zeta = 0$ (see Caldwell
\shortcite{ca} and Salopek \shortcite{sal}).

By looking at the definition (\ref{con}) of $\zeta$ we can strengthen the
previous result by showing that it must take non-zero values for growing modes
of the perturbations.
For such modes $\left|\Phi \right|$ has to increase with time since $\Phi$
characterizes the deviation from the unperturbed background metric. That
means that we must have
$\dot{\Phi} > 0$ for $\Phi > 0$ or $\dot{\Phi} < 0$ for $\Phi < 0$, at least
for some interval of time. Furthermore, the equation of state starts out
with $w$ close to $-1$ and then increases with time in realistic models of
inflation, such that $1+w \geq 0$. Since the Hubble parameter is always
positive, we see immediately from (\ref{con}) that $\zeta$ must take a
non-zero value during the time interval in question. But by its constancy
in the long-wavelength limit
this means that $\zeta \neq 0$ for all times for growing modes of metric
perturbations. The only chance for $\zeta$ to become zero is for decaying
modes where $\left|\Phi \right|$ decreases and $\Phi$ and $\dot{\Phi}$ have
opposite signs. But their contribution to the evolution of fluctuations
is negligible.

The situation when the equation of state is constant, e.g. after inflation
ended, requires a separate discussion, since a growing mode does not exist in
this case. Indeed, for $w = \rmn{const.}$ the differential equation for
metric fluctuations in a spatially flat universe has the general solution
\[
\Phi = c_1 + c_2 \exp\left( - \frac{5+3w}{2} \int H\, {\rmn{d}} t \right),
\]
where $c_1$ and $c_2$ are arbitrary constants, such that
the dominant mode is constant instead of growing. (The fastest way to
actually derive this solution is to use the conservation law (\ref{con}).)
But the conserved quantity becomes now
\[
\zeta = c_1 \left( 1 + \frac{2}{3} \frac{1}{1+w} \right),
\]
hence $\zeta$ is still non-zero if the dominant mode is present (i.e. if
$c_1 \neq 0$).

A simple, exactly solvable toy model will further exemplify the case with
inflation and varying $w$.

\section{The Toy Model}

\begin{figure}
\epsfxsize=3.3in \epsfbox{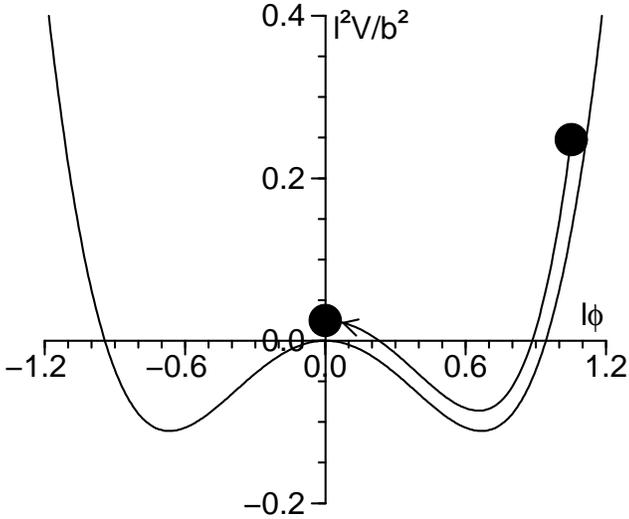}
\caption{The double well potential (\ref{pot}) plotted in rescaled units
as $l^2V/b^2$ versus $l\phi$. Initial conditions for the scalar field are
such that it asymptotically approaches the central maximum as
$t \rightarrow \infty$.}
\label{figpot}
\end{figure}

In the toy model, we use chaotic inflation with a simple, real scalar field
$\phi$ in a spatially flat universe with zero cosmological constant.
The potential is a double well of the form
\be
V(\phi) = b^2 \left[ \left(\frac{3l}{4}\right)^2\phi^4 -
\frac{1}{2}\phi^2 \right]
\label{pot}
\ee
with $l^2 = 8\upi G/3$ and $b>0$ an arbitrary parameter (see
Fig.~\ref{figpot}).
The time evolution of the scalar field during inflation
with this potential can not be solved for explicitly, unless we impose the
special initial conditions
\be
\phi(t_{\rmn{i}})=\phi_{\rmn{i}},\;
\dot{\phi}(t_{\rmn{i}})=\dot{\phi}_{\rmn{i}}
\stackrel{\textstyle !}{=}-b\phi_{\rmn{i}}
\label{initial}
\ee
at the beginning of inflation. Then we find (e.g. with Equ.~(6.9) and (6.10)
in Mukhanov et al. \shortcite{review})
\[
\phi(t) = \phi_{\rmn{i}} {\rmn{e}}^{-b\left( t-t_{\rmn{i}}\right) }
\]
and for the Hubble parameter
\be
H = \frac{3bl^2{\phi_{\rmn{i}}}^2}{4}
{\rmn{e}}^{-2b\left( t-t_{\rmn{i}}\right) },
\label{hubble}
\ee
both simple exponential decays.

It should be pointed out that initial conditions (\ref{initial}) involve
serious fine-tuning as we give the scalar field just enough kinetic energy
to asymptotically roll up the central maximum of the potential. In this
toy model, the potential will not end up in one of the minima of the
potential.

The equation of state evolves as
\[
w = {\left( \frac{4}{3l\phi_{\rmn{i}}} \right)}^2
    {\rmn{e}}^{2b\left( t-t_{\rmn{i}}\right)} - 1,
\]
so in order to get a sufficient amount of inflation at early times we have to
require
\be
{\left( \frac{3l\phi_{\rmn{i}}}{4} \right)}^2 \gg 1.
\label{slow}
\ee
Using the Hubble parameter (\ref{hubble}) we see that this condition is
equivalent to $1/b \gg 1/H$ for times close to $t_{\rmn{i}}$, i.e. the time
scale of change of the scalar field is much larger than the Hubble time.
Thus (\ref{slow}) is just a slow rolling condition for the initial evolution
of the scalar field.

Note though that in this scenario $w$ does not approach a
finite value, like $0$ or $1/3$, but keeps growing indefinitely. This is caused
by both the pressure and energy density of the scalar field approaching
zero at late times, but in such a way that their ratio tends to infinity.
In a realistic model of inflation we should also consider the presence of
matter and/or radiation which will dominate pressure and energy density at
late times when the contribution from the scalar field has fallen off to low
values, with $w$ approaching the level typical for matter and/or radiation.
Hence the indefinite growth of the equation of state in this model just
indicates that it is useful only up to a certain time. The slow rolling
condition (\ref{slow}) ensures that the model is applicable long enough to
get a sufficient amount of inflation.

\section{Metric Perturbations in the Toy Model}

\begin{figure}
\epsfxsize=3.3in \epsfbox{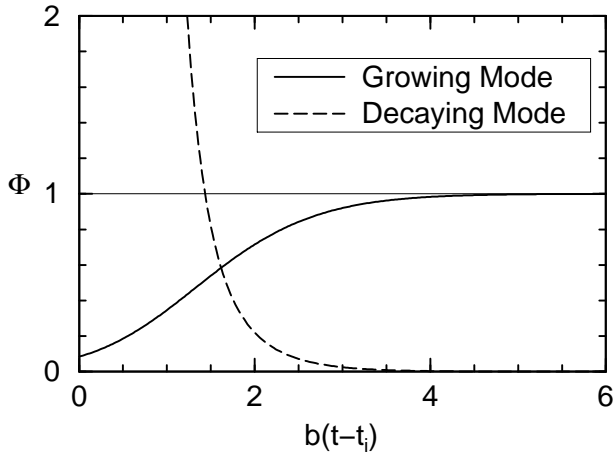}
\caption{Time evolution of gauge-invariant metric perturbations $\Phi$ for
the decaying mode (\ref{decay}) and the growing mode (\ref{grow}) as a
function of $b\left( t-t_{\rmn{i}}\right)$. $3l^2{\phi_{\rmn{i}}}^2/8 = 10$
has been chosen for this plot. The decaying mode is finite at
$t = t_{\rmn{i}}$
and tends to infinity as $t \rightarrow -\infty$, whereas the growing mode
approaches zero for $t \rightarrow -\infty$.}
\label{figphi}
\end{figure}

The time evolution of the gauge-invariant generalized Newtonian potential
$\Phi$, which characterizes scalar metric perturbations, is governed by
Equ.~(\ref{phiequ}). It is hard to solve in general, but proper substitutions
reduce this differential equation to a confluent hypergeometric one in our
case. That way, two linearly independent base solutions in the
long-wavelength limit are found as
(using the relations and series expansions for confluent hypergeometric
functions from Section~(13.1) in Abramowitz \& Stegun \shortcite{as})
\be
\Phi = \tau {\rmn{e}}^\tau
\label{decay}
\ee
and
\be
\Phi = 1 + \tau \left[ {\rmn{e}}^\tau \left( \gamma + \ln \tau \right) -
       \sum_{k=1}^{\infty} \frac{\tau^k}{k!}
       \sum_{\nu=1}^{k} \frac{1}{\nu} \right]
\label{grow}
\ee
with the abbreviation
\[
\tau = \frac{3l^2{\phi_{\rmn{i}}}^2}{8}
{\rmn{e}}^{-2b\left( t-t_{\rmn{i}}\right) }
\]
and Euler's constant $\gamma$ (see Fig.~\ref{figphi}). The general solution
is then a linear combination of (\ref{decay}) and (\ref{grow}).

Clearly, (\ref{decay}) is a decaying mode, and therefore of not much
importance in the evolution of fluctuations, whereas a discussion of
(\ref{grow}) shows that $\Phi > 0$ and $\dot{\Phi} > 0$ for this base
solution, hence it is a growing mode. Taking the two modes
(\ref{decay}) and (\ref{grow}), a straightforward calculation of the
conserved quantity (\ref{con}) then shows $\zeta=0$ for the decaying mode,
but $\zeta=1$ for the growing one.

\section{Conclusions}

The `conservation law' (\ref{con}) for cosmological perturbations indeed is
a useful relation describing the growth of fluctuations of the metric in the
long-wavelength limit outside the horizon, and it is derived from the
correct equations of motion. It has been shown that the
important growing modes always yield a non-zero value for the conserved
quantity $\zeta$, both on general grounds and in an explicitly solvable
toy model. The situation $\zeta=0$, when (\ref{con}) can not be used to make
any statement on the growth of metric perturbations in the long-wavelength
limit, can occur only for decaying modes, which do not play a significant
r\^{o}le in the evolution of cosmological perturbations.

After completion of this work a preprint appeared \cite{martin} in which
similar conclusions concerning the validity of Grishchuk \shortcite{gr} are
reached.

\section*{Acknowledgments}

I wish to thank Robert Brandenberger for discussions and comments
on the first draft of this paper and L. P. Grishchuk for useful
communication.

\end{document}